\documentclass[12pt]{spieman}  %
\usepackage{amsmath,amsfonts,amssymb}
\usepackage{graphicx}
\usepackage{setspace}
\usepackage{tocloft}
\usepackage{tabularx}
\usepackage{booktabs}
\usepackage{acronym}

\acrodef{US}{ultrasound}
\acrodef{OptA}{optoacoustic}
\acrodef{OPUS}{optoacoustic-ultrasound}
\acrodef{MSOPUS}[MS-OPUS]{multispectral optoacoustic-ultrasound}
\acrodef{ROI}{region of interest}
\acrodefplural{ROI}{regions of interest}
\acrodef{GUI}{graphical user interface}
\acrodef{PAD}{peripheral artery disease}
\acrodef{AUROC}{area under the receiver operating characteristic curve}
\acrodef{PATATO}{Photoacoustic Tomography Analysis Toolkit}
\acrodef{IPASC}{International Photoacoustic Standardisation Consortium}
\acrodef{BP}{backprojection}
\acrodef{LU}{linear spectral unmixing}

\title{OPTARI: An open-source software framework for clinical translation of optoacoustic imaging}

\author[a,*]{Moritz Schillinger}
\author[b,c]{Maximilian Bader}
\author[d]{Adrian Buehler}
\author[e]{Felix Wachter}
\author[e]{Ferdinand Knieling}
\author[b,c,f]{Vasilis Ntziachristos}
\author[a,g]{Katharina Breininger}
\affil[a]{Department Artificial Intelligence in Biomedical Engineering (AIBE), Friedrich-Alexander-Universität Erlangen-Nürnberg, Erlangen, Germany}
\affil[b]{Chair of Biological Imaging, Central Institute for Translational Cancer Research (TranslaTUM), School of Medicine and Health \& School of Computation, Information and Technology, Technical University of Munich, Munich, Germany}
\affil[c]{Institute of Biological and Medical Imaging, Bioengineering Center, Helmholtz Zentrum München, Neuherberg, Germany}
\affil[d]{Department of Vascular Surgery, University Hospital Erlangen, Friedrich-Alexander-Universität Erlangen-Nürnberg, Erlangen, Germany}
\affil[e]{Department of Pediatrics and Adolescent Medicine, University Hospital Erlangen, Friedrich-Alexander-Universität Erlangen-Nürnberg, Erlangen, Germany}
\affil[f]{Munich Institute of Robotics and Machine Intelligence (MIRMI), Technical University of Munich, Munich, Germany}
\affil[g]{Center for AI and Data Science (CAIDAS), Julius-Maximilians-Universität Würzburg, Würzburg, Germany}

\cftpagenumbersoff{figure}
\cftpagenumbersoff{table} 
\begin{document} 
\maketitle

\begin{abstract}

\noindent \textbf{Significance}: Combined \ac{OptA} and ultrasound imaging is a promising modality across a wide range of pathologies. Its use in clinical research, however, is limited by a lack of analysis software accessible to users without programming expertise. Existing tools are either script-based engineering frameworks or proprietary vendor applications.

\noindent \textbf{Aim}: We present OPTARI, an open-source graphical software application to facilitate large-scale clinical \ac{OPUS} studies.

\noindent \textbf{Approach}: OPTARI builds on the napari image viewer and the PATATO toolkit for \ac{OptA} data processing. It supports native vendor and open HDF5 data formats, aligns with the IPASC standard, and is distributed as a standalone executable. Features include region-of-interest annotation, reconstruction, chromophore unmixing, deep-learning-based tissue segmentation, and temporal and spectral analysis modules.

\noindent \textbf{Results}: OPTARI offers a combination of features for clinical study analysis that none of the compared tools provide. The automated tissue segmentation produced results comparable to expert labeling for muscle tissue, and batch processing reduced analysis time from minutes to under one second per scan. The tool is used in ongoing clinical \ac{OPUS} studies.

\noindent \textbf{Conclusions}: OPTARI connects engineering frameworks with medical research and provides an open-source basis for reproducible and standardized analysis in clinical \ac{OptA} imaging studies.

\end{abstract}

\acresetall


{\noindent \footnotesize\textbf{*}Corresponding author: Moritz Schillinger, \linkable{moritz.schillinger@fau.de} }

\begin{spacing}{1}   %

\section{Introduction}
\label{sect:intro} 

By combining the strengths of both \ac{OptA} and \ac{US} imaging, dual-modal \ac{OPUS} imaging has shown great potential across a wide range of clinical applications~\cite{Park2024-review, Ntziachristos2024-clinicalNeed}. While conventional \ac{US} visualizes the mechanical and structural properties of tissue, \ac{OptA} imaging provides functional and molecular information, such as blood oxygenation and chromophore concentrations~\cite{Beard2011-review}. Since \ac{OptA} and \ac{US} share the same sensor geometry and acoustic detection hardware, dual-modal scanners can acquire both images simultaneously, generating inherently co-registered datasets with minimal acquisition overhead.

Extracting quantitative information from these datasets requires dedicated software for reconstruction, spectral unmixing, and analysis. \Ac{OptA} signal processing has advanced substantially in recent years~\cite{Chowdhury2021-sTIR, Noelke2024-cinn, Bader2024-SPR, Dehner2025-Deepmb2}. Yet, despite this engineering progress, clinical research studies rarely make use of the latest algorithms, largely because researchers outside of engineering lack a straightforward way to access them. Based on recent clinical \ac{OptA} reviews~\cite{Park2024-review, Knieling2025-primer} and feedback from our clinical collaborators, we identified three challenges currently limiting clinical research in \ac{OptA} on the software and data-processing side:

\begin{enumerate}
    \item \textbf{Data Standardization and Interoperability}: Clinical \ac{OptA} scans are typically saved in proprietary, manufacturer-specific formats. This fragmentation makes it difficult to migrate older datasets, share data between research centers, or perform custom downstream analysis. The \ac{IPASC} has proposed unified, open formats to store and share \ac{OptA} imaging data. However, this has not yet seen widespread adoption in clinical studies~\cite{Groehl2022-ipasc, Assi2023-roadmap}
    \item \textbf{Workflow Subjectivity and Operator Variability}: The quantitative analysis of \ac{OptA} data involves tuning a long list of parameters. Even a standard clinical analysis requires choosing a reconstruction algorithm, defining unmixing spectra, choosing the right frame, and selecting \acp{ROI} to extract chromophore concentrations. Many of these steps depend on subjective expert judgment. This manual approach introduces intra- and inter-operator variability. To make \ac{OPUS} studies reproducible and comparable, it is important to move from user-dependent steps to objective workflows.
    \item \textbf{Technical Barriers and Parameter Abstraction}: Large clinical studies need fast, intuitive analysis, but manually configuring complex reconstruction and unmixing parameters is impractical in this setting, so clinical users benefit from well-documented default settings that can be saved and shared easily. Many studies also do not need the full content of research scans, such as raw time series or long frame sequences, so analysis tools should abstract this complexity and focus on the relevant data and metrics.
\end{enumerate}

Existing software falls into two groups, each with its own limitations. Proprietary vendor software, like iLabs (iThera Medical, Munich, Germany) or Imagio/SenoGram (Seno Medical Instruments, San Antonio, TX, USA), is typically locked to specific hardware or data formats, lacks capacity for custom extensions, and depends on the manufacturer for maintenance. In contrast, a small number of open-source frameworks exist for clinical \ac{OptA} data, ranging from simulation-focused toolkits~\cite{Groehl2022-simpa} to script-based reconstruction and unmixing libraries with limited annotation support~\cite{Else2024-patato}. However, none of these open-source tools combine a graphical interface with the structured, multi-patient workflow required for large-scale clinical studies.

To bridge this gap between engineering frameworks and clinical \ac{OPUS} studies, we introduce OPTARI (\textbf{OPT}oacoustic imaging toolkit based on nap\textbf{ARI}), an open-source and platform-agnostic software application\footnote{\linkable{https://mo-sc.github.io/OPTARI}}. Rather than reimplementing existing functionality, OPTARI extends established open-source tools, like the \ac{PATATO} for core data management~\cite{Else2024-patato} and napari for the graphical interface~\cite{Chiu2022-napari}, and combines them with an analysis framework optimized for a standard clinical \ac{OPUS} study workflow. Deterministic, shareable configurations are included to reduce operator variability, while automated \ac{ROI} placement and batch processing reduce the time required to process large datasets. OPTARI does not define standards or specific parameters itself. Instead, it lets clinical researchers adjust processing configurations, test them, and share them with collaborators, which may help the field to establish common defaults.

The next section defines the scope of OPTARI and provides an overview of its core features. Section~\ref{sect:architecture} outlines the underlying software architecture and implementation details. Section~\ref{sect:evaluation} compares OPTARI to related tools, describes its use in clinical studies, and evaluates the automated analysis in terms of segmentation performance and processing time, followed by discussion and limitations.

\section{Scope and Features}\label{sect:scope}

OPTARI focuses specifically on clinical two-dimensional handheld \ac{MSOPUS} imaging. This setup forms the basis for the majority of clinical studies across manufacturers, including application in inflammatory bowel disease~\cite{Knieling2017-crohn, Regensburger2024-ibd}, breast cancer~\cite{Neuschler2018-cancer, Menezes2018-cancer}, muscular disorders~\cite{Regensburger2019-dmd, Regensburger2022-sma}, and cardiovascular conditions~\cite{Traeger2024-pad, Caranovic2025-pad}.

\subsection{System Requirements}

We formulated six requirements that follow directly from the challenges identified in Section~\ref{sect:intro}, summarized in Table~\ref{tab:requirements}: data standardization (R6), operator variability (R4), and technical barriers (R1, R2, R3, R5).

\begin{table}[htbp]
\centering
\caption{System requirements for OPTARI.}
\label{tab:requirements}
\begin{tabularx}{\linewidth}{l l X}
\toprule
 & \textbf{Requirement} & \textbf{Description} \\
\midrule
\textbf{R1} & Easy Deployment & Installation and execution require no command-line interaction or script execution. Distribution as a standalone executable that runs without administrative privileges, so it can be deployed in restricted hospital IT environments. \\
\addlinespace
\textbf{R2} & Graphical Interface & Provide a click-based \acf{GUI} for intuitive navigation, paired with keyboard shortcuts to speed up repetitive workflows. \\
\addlinespace
\textbf{R3} & Study Navigation & Support analysis across multiple subjects and scans, easy switching between scans, and fast scrolling through temporal and spectral layers. \\
\addlinespace
\textbf{R4} & Objective Processing & Replace subjective parameter selection with deterministic procedures as much as reasonably possible, and allow easy sharing of parameters and configurations. \\
\addlinespace
\textbf{R5} & Documentation & Provide feature documentation and step-by-step guides for common clinical study workflows. \\
\addlinespace
\textbf{R6} & Interoperability & Read and write both vendor-native and open data formats, and keep the underlying code accessible and documented so that custom modules or data formats can be integrated. \\
\bottomrule
\end{tabularx}
\end{table}

\subsection{Feature Overview}

\begin{figure*}[htbp]
  \centering
  \includegraphics[width=\textwidth]{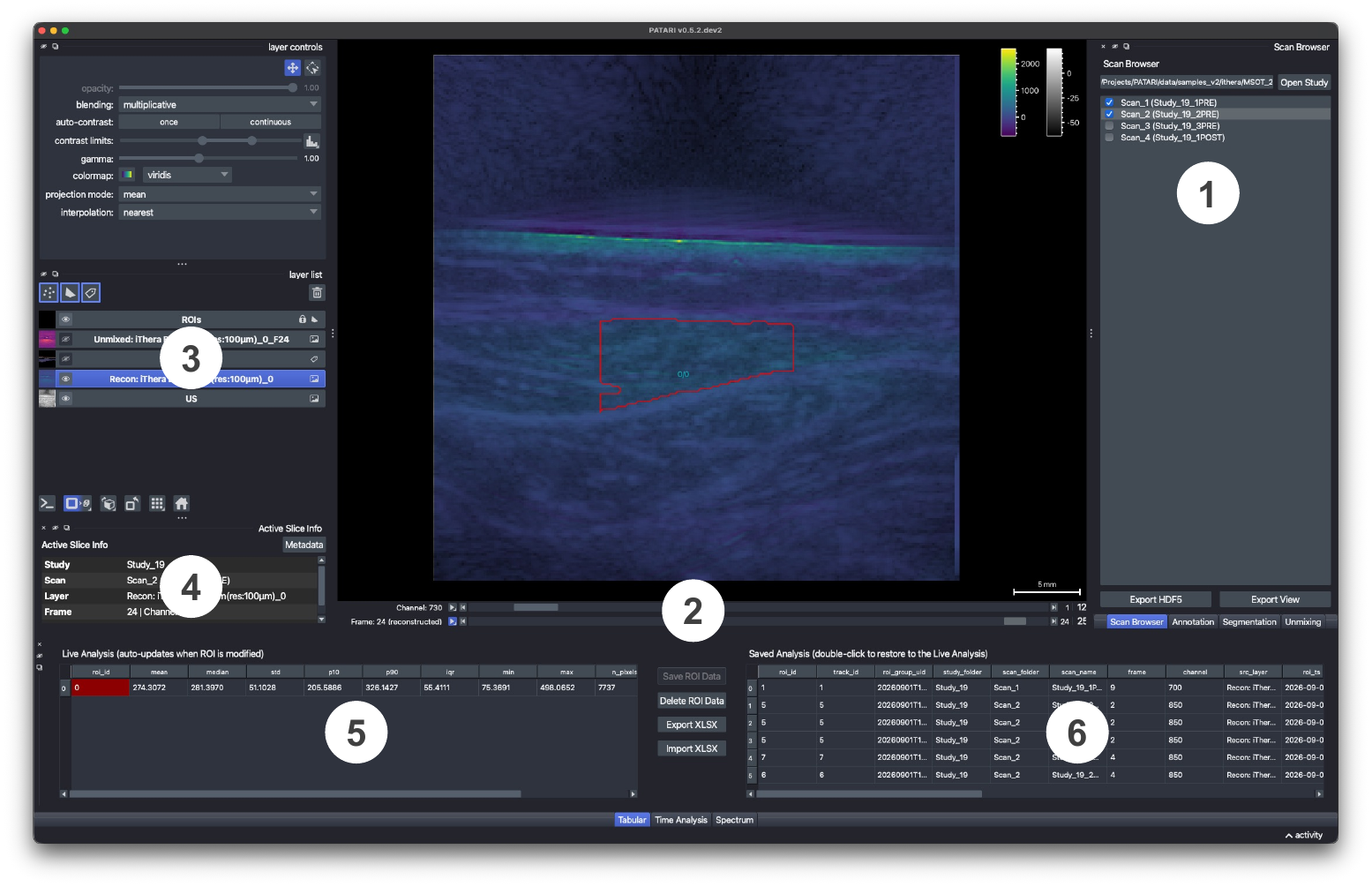}
  \caption{\label{fig:patari-screenshot-main} Main viewer window of OPTARI. (1) \textit{Scan Browser} for study navigation. (2) Sliders for temporal and spectral scrolling. (3) \textit{Layer List}, showing image data and annotations. (4) Metadata viewer. (5) \textit{Live Analysis} table, showing dynamic updates of \ac{ROI} data. (6) \textit{Saved Analysis} table to collect data across scans.}
\end{figure*}

The following sections break down the main features of OPTARI. Figures~\ref{fig:patari-screenshot-main} and \ref{fig:patari-screenshot-widget} show screenshots of the user interface (OPTARI v0.8.0), with numbered labels matching the specific features described in the text.

\subsubsection{Viewing and Navigation}

OPTARI's interface uses a \textit{Scan Browser} that makes navigating multi-scan studies fast and straightforward (1, Fig.~\ref{fig:patari-screenshot-main}). Sliders give users direct control over multi-dimensional image stacks, making it easy to scroll through wavelengths, frames, and unmixed tissue components (2, Fig.~\ref{fig:patari-screenshot-main}). A motion-scoring algorithm automatically selects the frame with the least motion in the sequence~\cite{Kukacka2026-motionscore}. The main viewer displays \ac{US}, \ac{OptA}, and annotation layers in a flexible \textit{Layer List} (3, Fig.~\ref{fig:patari-screenshot-main}). Users can adjust opacity, contrast limits, and colormaps to overlay \ac{OptA} data onto \ac{US} images. Available metadata, such as timestamps and acquisition settings, are displayed alongside the image using \ac{IPASC}-compliant tag names and units (4, Fig.~\ref{fig:patari-screenshot-main}).

\subsubsection{Annotation, ROI Management, and Automated Segmentation}

For quantitative evaluation, users can draw or edit custom and predefined \acp{ROI}, like polygons and ellipses, directly on the image. OPTARI automatically extracts customizable statistical features from these regions, which can correspond to chromophore concentrations or derived parameters like tissue oxygenation. Data collection relies on a dual-table system: the \textit{Live Analysis} table dynamically updates measurements whenever an \ac{ROI} is moved, resized, or switched to another frame (5, Fig.~\ref{fig:patari-screenshot-main}), while the \textit{Saved Analysis} table freezes selected data points into a permanent log (6, Fig.~\ref{fig:patari-screenshot-main}). 

A deep-learning-based tissue \textit{Segmentation} module can segment key tissues for fully automatic \ac{ROI} placement (7, Fig.~\ref{fig:patari-screenshot-widget}), with pretrained models provided for common examination sites in clinical \ac{OPUS} studies. Users can also save custom \ac{ROI} presets to an \textit{ROI Library}, backed by a JSON file, making it easy to share \ac{ROI} configurations across team members or workstations (8, Fig.~\ref{fig:patari-screenshot-widget}). These presets scale to each scan's field of view and can be placed at their saved coordinates or anchored automatically onto a matching tissue segmentation class. Applying a preset across multiple frames groups the \acp{ROI} into a single track (a set of \acp{ROI} linked as one region across frames), where each frame is anchored independently on its own segmentation mask, so the track follows the target tissue.

\subsubsection{Temporal and Spectral Analysis}

OPTARI provides interactive plotting tools to help inspect temporal dynamics and spectral properties directly within the \ac{GUI} for exploratory data analysis. The \textit{Temporal Analysis} module uses timestamps from the scan metadata to plot intensities across groups of \acp{ROI}, enabling tracking of chromophore concentrations or tissue oxygenation over time, even in scans with motion artifacts (9, Fig.~\ref{fig:patari-screenshot-widget}). The \textit{Spectral Analysis} module displays signal intensities as a function of wavelength to show the spectral distribution (10, Fig.~\ref{fig:patari-screenshot-widget}).

\subsubsection{Data Processing and Conversion}

In addition to the \ac{PATATO} data loaders for HDF5 files and MSOT Acuity scans (iThera Medical GmbH, Munich, Germany), we implemented support for the newer MSOT Frontier scanner (iTheraNEXT Imaging GmbH, Munich, Germany). 

\begin{figure}[htbp]
  \centering
  \includegraphics[width=\textwidth]{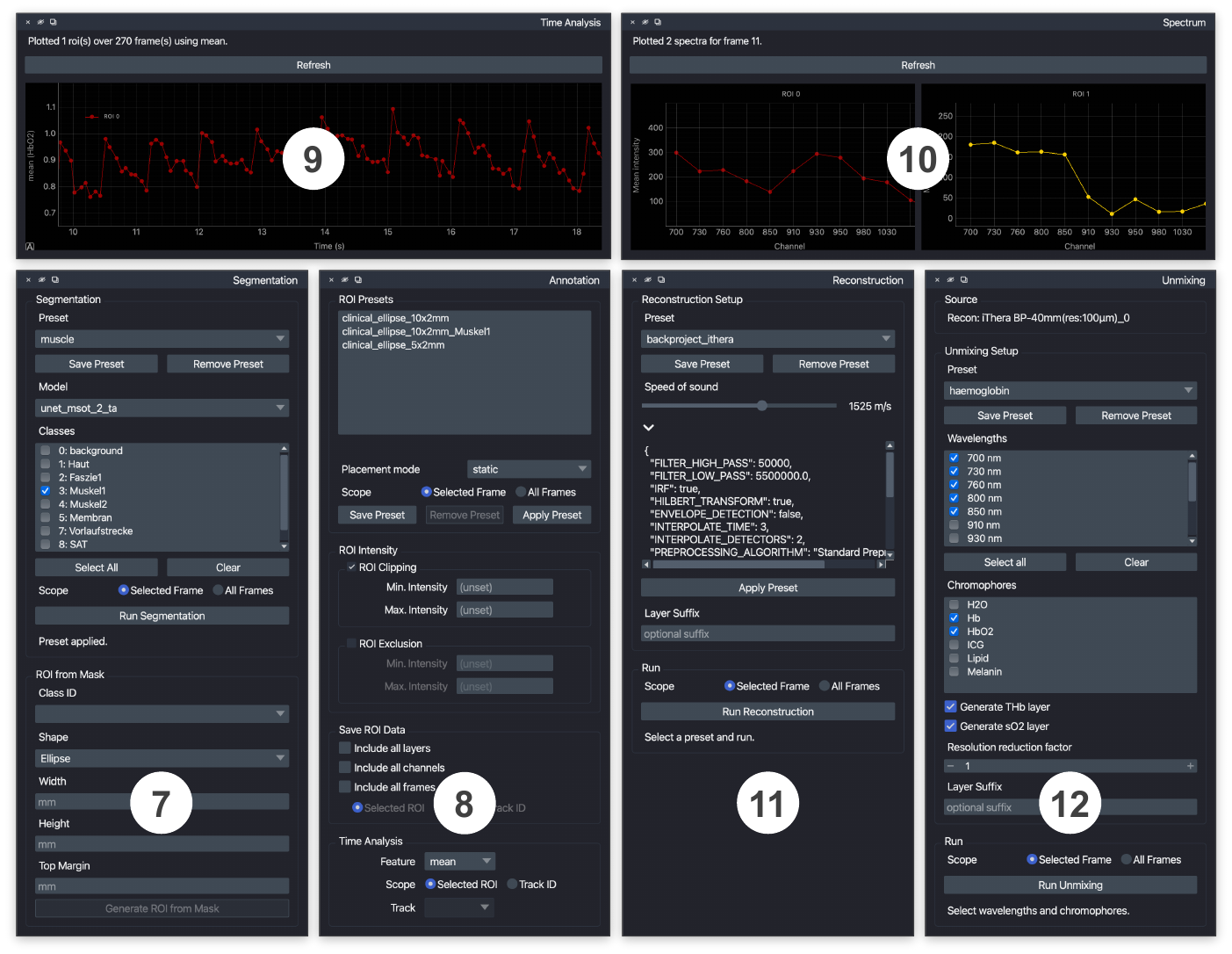}
  \caption{\label{fig:patari-screenshot-widget} Selection of processing and analysis widgets in OPTARI. (7) \textit{Segmentation} module to automatically place an \ac{ROI} in the target tissue. (8) \textit{ROI Library} for loading and saving \ac{ROI} presets. (9) \textit{Temporal Analysis} module to plot signal intensity changes over time. (10) \textit{Spectral Analysis} module to visualize signal magnitude across wavelengths. (11) \textit{Reconstruction} module for \ac{BP}- and deep-learning-based image reconstruction. (12) \textit{Unmixing} module for spectral unmixing of chromophore concentrations.}
\end{figure}

OPTARI wraps \ac{PATATO}’s processing features into \ac{GUI} widgets so users can run \textit{Reconstruction} (11, Fig.~\ref{fig:patari-screenshot-widget}) and \textit{Unmixing} (12, Fig.~\ref{fig:patari-screenshot-widget}). Reconstruction settings include signal preprocessing steps such as band-pass filtering and impulse response correction. Results are automatically added to the viewer as interactive layers. In addition to the included \acf{BP}-based reconstruction algorithms, we added an adapter for DeepMB, a state-of-the-art deep-learning model trained on data from MSOT Acuity scanners~\cite{Dehner2023-deepmb}. Chromophore concentrations are calculated using \ac{LU}, with reference spectra available for water ($\text{H}_2\text{O}$), oxygenated hemoglobin ($\text{HbO}_2$), deoxygenated hemoglobin ($\text{Hb}$), indocyanine green (ICG), lipid, and melanin~\cite{Prahl2018-spectra}. These can also be used to calculate derived features like oxygen saturation ($\text{SO}_2$) or total hemoglobin (THb). Processing parameters can be loaded from and stored in JSON files to re-use or share them with collaborators. 

\subsubsection{Data Export}

Exporting data in OPTARI automatically wraps processed scans, unmixed layers, drawn \acp{ROI}, and segmentation masks into HDF5 files. This provides an easy way to convert proprietary vendor data into open HDF5 files that include \ac{IPASC}-compliant metadata. All extracted metrics and metadata from the \textit{Saved Analysis} table can be exported with a single click into standard spreadsheet formats, which can also be re-imported to resume an analysis or combine results from several sessions or machines. Analytical plots like histograms or spectral curves can be exported directly as image files. Additionally, the current viewer contents can be saved as figures or compiled across frame sequences into video exports.

\subsubsection{Batch Mode}\label{sect:batchmode}

\textit{Batch Mode} runs a previously defined pipeline of reconstruction, unmixing, segmentation, and \ac{ROI} placement across an entire dataset without manual interaction. Each step is tuned once on a representative scan and saved as a preset. A batch preset then references these presets and the full analysis plan is validated so that incompatibilities or missing requirements are flagged before the run starts. Individual scan failures are logged and skipped. The output is a measurement table with the same structure as the \textit{Saved Analysis} table in the \ac{GUI}, along with optional overlay images and HDF5 conversion.

\section{Implementation and Architecture}\label{sect:architecture}

This section describes how OPTARI is built, some of the external tools we integrated, and how it is deployed.

\subsection{Software Architecture}

\begin{figure*}[htbp]
  \centering
  \includegraphics[width=\textwidth]{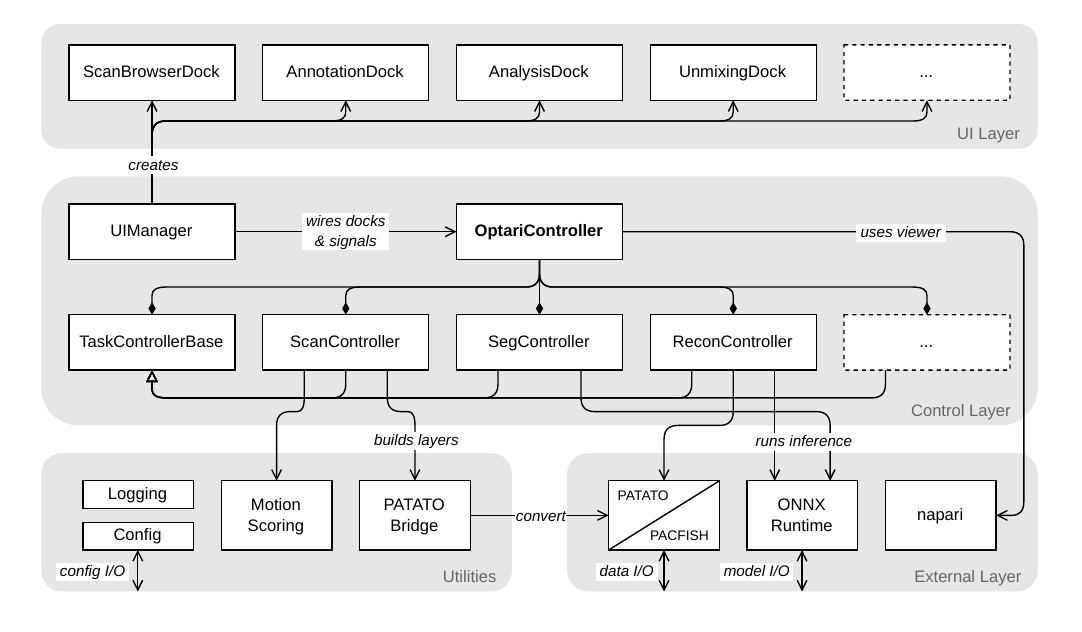}
  \caption{\label{fig:patari-diagram} Overview of the modular software architecture and selected components of OPTARI: UI Layer (dock widgets), Control Layer (central OptariController and specialized task modules), Utilities Layer, and External Layer.}
\end{figure*}

OPTARI uses a Model-View-Controller (MVC)-inspired architecture to separate the graphical interface from the data processing logic. As shown in Fig.~\ref{fig:patari-diagram}, the application is divided into four main layers: a UI Layer, a Control Layer, a Utilities Layer, and an External Layer. Rather than running as a standard napari plugin, OPTARI embeds the napari viewer directly into a custom Qt\footnote{\linkable{https://doc.qt.io/qtforpython-6}; v6.11.2} application. The UI Layer contains only graphical widgets that hold no internal logic. Instead, a stateless \textit{UIManager} builds these widgets and wires UI signals to backend routines, while a separate \textit{ShortcutManager} handles custom keyboard commands.

The core of the application is the \textit{OptariController}. It manages global application state, like tracking open widgets, the active study, and the loaded scan, while delegating specific tasks to modular task controllers. For example, the \ac{ROI} annotation subsystem uses a dedicated controller to monitor napari's shape layers. Whenever a user draws or resizes an \ac{ROI}, the controller automatically recalculates tissue metrics and refreshes the live data tables. The \textit{SegController} manages tissue segmentation by running inference via pre-trained deep learning models. Based on the generated masks, \acp{ROI} can be placed automatically in the target tissue and added to the analysis tables. We include model weights pre-trained or fine-tuned on US images from three examination sites common in clinical \ac{OPUS} studies: the gastrocnemius muscle, the tibialis anterior muscle, and the vastus lateralis muscle~\cite{Knieling2025-primer}. Model weights are automatically downloaded during their first execution to keep the base installation size small. New models can easily be plugged in via the built-in ONNX adapter. Other task controllers wrap or extend processing algorithms provided by external frameworks and automatically convert the external outputs into interactive napari layers. Long-running operations such as reconstruction, segmentation, and unmixing run in a background task slot so the event loop stays responsive while a scan processes. The same slot is also used by \textit{Batch Mode} (Section~\ref{sect:batchmode}).

For real-time data inspection, we used pyqtgraph\footnote{\linkable{https://github.com/pyqtgraph/pyqtgraph}; v0.14.0} to build analysis widgets docked below the viewer that can dynamically plot intensity distributions, wavelength spectra, or temporal signal changes in the \acp{ROI}.

\subsection{External Integrations}

OPTARI integrates several external tools to handle data I/O, image reconstruction, and quality control. For data management, a custom patato-bridge utility connects the software to an extended version of the \ac{PATATO} framework (based on v0.7.0)~\cite{Else2024-patato}, which natively handles both proprietary vendor files (e.g., iThera .msot datasets) and standardized HDF5 datasets. Through this integration, users can also run \ac{PATATO}’s JAX-accelerated \ac{BP} algorithm. We choose a \ac{PATATO}-inspired HDF5 schema over the format provided directly by the \ac{IPASC}, because the native \ac{IPASC} format is scoped to raw time-series data and does not accommodate dual-modality ultrasound or processed/reconstructed layers~\cite{Groehl2022-ipasc}. Nevertheless, OPTARI stores \ac{IPASC}'s consensus metadata tags alongside its own HDF5 schema. Native \ac{IPASC} files can be imported and exported directly from OPTARI using PACFISH\footnote{\linkable{https://github.com/IPASC/PACFISH}; v0.4.4}, which also performs metadata consistency checks.

Out of the increasing number of deep-learning-based \ac{OptA} image reconstruction algorithms, none has been widely adopted in published clinical \ac{OPUS} studies. Apart from general concerns over black-box predictions in \ac{OptA} imaging~\cite{Knopp2025-bias}, this is largely a problem of training on simulated data or non-standard scanner geometries. To our knowledge, DeepMB is currently the only deep-learning reconstruction framework trained specifically on clinical \ac{OptA} data recorded with the widely deployed MSOT Acuity scanner and has been integrated as the standard reconstruction software in the latest MSOT Frontier scanner~\cite{Dehner2025-Deepmb2}. It achieved near-identical image quality to iterative model-based algorithms at an approximately 1,000-fold acceleration~\cite{Dehner2023-deepmb}. Like the segmentation models, DeepMB runs on the ONNX Runtime within OPTARI. This balances performance with hardware compatibility, since it uses GPU acceleration where available and falls back to CPU execution if the required hardware or drivers are not present.

To reduce operator bias during frame selection, OPTARI incorporates a motion scoring algorithm that quantifies movement in sequences of \ac{US} frames based on their structural similarity index and zero-mean normalized cross-correlation, adapted from the frame-selection approach of Kukačka et al.~\cite{Kukacka2026-motionscore}. The software then automatically suggests the frame with the lowest motion score for subsequent analysis.

\subsection{Deployment and Installation}

To simplify deployment in hospital environments, OPTARI is packaged as a standalone desktop application using PyApp\footnote{\linkable{https://github.com/ofek/pyapp}; v0.29.0}. PyApp wraps the project into a single launcher executable that sets up an isolated runtime on its first launch and requires no administrative rights or terminal commands. Because certain \ac{PATATO} modules depend on native C/C++ code, we host and distribute precompiled \ac{PATATO} binaries for Windows, macOS, and Linux (Ubuntu), removing any requirement for local compilers. We use an automated GitHub Actions CI/CD pipeline to handle unit testing, tag-based versioning, cross-platform release builds and documentation builds. OPTARI can be customized through a \ac{GUI} dialog that covers viewer defaults, dock visibility, and \ac{ROI} table display. The presets for reconstruction, segmentation, unmixing, and \ac{ROI} placement are managed separately from their own dock. All settings and presets are stored as human-readable JSON files in the user's home directory.

\section{Evaluation}\label{sect:evaluation}

We compared OPTARI to existing software, described how it was integrated into clinical study workflows, and evaluated the performance of its automated segmentation and \ac{ROI} placement.

\subsection{Feature Comparison}\label{sect:feature_comparison}

We included only tools that offer some form of graphical access to \ac{OptA} data processing (Table~\ref{tab:feature_comparison}). iLabs, iThera Medical's proprietary clinical analysis software for \ac{OptA} data, offers a mature interface with clinical documentation, but remains closed-source and locked to iThera's own hardware and data formats. \Ac{PATATO} is currently the closest open-source tool to clinical utility, offering reconstruction and unmixing capabilities alongside minimal \ac{GUI} and annotation features~\cite{Else2024-patato}, but remains largely script-based and lacks a structured, multi-patient study workflow. The MITK platform offers an \ac{OptA} plugin that integrates \ac{OptA} data visualization~\cite{Kirchner2019-mitkpai} but it misses important features like spectral unmixing or advanced clinical annotation tools, and only provides limited data format support. PATLAB offers a \ac{GUI}-based reconstruction and simulation environment aimed at system and algorithm development, but lacks spectral unmixing, clinical annotation tools, a multi-patient workflow, and requires a MATLAB license~\cite{Omidi2022-patlab}. ImageJ/Fiji is a general-purpose, extensible image analysis platform and is included as a common fallback tool clinical researchers use in the absence of dedicated software~\cite{Schindelin2012-fiji}. However, it does not offer any \ac{OptA}-specific reconstruction, unmixing, or annotation features.

\begin{table*}[t]
\centering
\caption{Feature comparison of OPTARI against existing software tools used in clinical \ac{OptA} imaging. Evaluation criteria correspond to the requirements (R1--R6) defined in Section~\ref{sect:scope}. Assessment based on the software versions and documentation available in September 2026.}
\label{tab:feature_comparison}
\footnotesize
\setlength{\tabcolsep}{4pt}
\begin{tabular}{lcccccc}
\toprule
\textbf{Requirement / Feature} & \textbf{iLabs} & \textbf{\ac{PATATO}} & \textbf{MITK-PAI} & \textbf{PATLAB} & \textbf{ImageJ/Fiji} & \textbf{OPTARI} \\
\midrule
\textbf{R1: Easy Deployment} & & & & & & \\
\quad -- Standalone executable & $+$ & $-$ & $+$ & $-$ & $+$ & $+$ \\
\quad -- No admin rights required & $+$ & $\circ$ & $\circ$ & $-$ & $+$ & $+$ \\
\midrule
\textbf{R2: Graphical Interface} & & & & & & \\
\quad -- \ac{GUI}-driven analysis & $+$ & $\circ$ & $+$ & $\circ$ & $+$ & $+$ \\
\quad -- Keyboard shortcuts & $-$ & $-$ & $\circ$ & $-$ & $+$ & $+$ \\
\midrule
\textbf{R3: Study Navigation} & & & & & & \\
\quad -- Multi-subject workflow & $+$ & $-$ & $-$ & $-$ & $-$ & $+$ \\
\quad -- Temporal/spectral scrolling & $+$ & $\circ$ & $\circ$ & $\circ$ & $+$ & $+$ \\
\midrule
\textbf{R4: Objective Processing} & & & & & & \\
\quad -- Shareable parameters & $\circ$ & $+$ & $-$ & $\circ$ & $-$ & $+$ \\
\quad -- Automated \ac{ROI} placement & $-$ & $-$ & $-$ & $-$ & $-$ & $+$ \\
\midrule
\textbf{R5: Documentation} & & & & & & \\
\quad -- Clinical \ac{OptA} documentation & $+$ & $\circ$ & $-$ & $-$ & $-$ & $+$ \\
\quad -- Code documentation & $-$ & $+$ & $+$ & $+$ & $+$ & $+$ \\
\midrule
\textbf{R6: Interoperability} & & & & & & \\
\quad -- Vendor \& open data support & $\circ$ & $+$ & $\circ$ & $\circ$ & $\circ$ & $+$ \\
\quad -- Open-source codebase & $-$ & $+$ & $+$ & $+$ & $+$ & $+$ \\
\midrule
\multicolumn{7}{l}{\footnotesize $+$ fully supported \quad $\circ$ partially supported \quad $-$ not supported} \\
\bottomrule
\end{tabular}
\end{table*}

Beyond these, several other tools exist outside the scope of this comparison. SIMPA (Simulation and Image Processing for Photonics and Acoustics) is one of the most widely used frameworks for \ac{OptA} simulation and covers the full pipeline from synthetic data generation to reconstruction~\cite{Groehl2022-simpa}. However, it does not offer a \ac{GUI}, annotation, or clinical workflow features. 
PHANTOM is a MATLAB tool for ultrasound-guided fluence compensation in preclinical \ac{OptA} imaging, not a general clinical analysis platform~\cite{Sweeney2025-phantom}. MSOT-RAFT is a script-based reconstruction and filtering toolbox for preclinical \ac{OptA} data with no graphical interface~\cite{OKelly2021-msotraft}.

\subsection{Integration into Clinical Studies}

OPTARI was developed in close collaboration with clinical professionals experienced in \ac{OPUS} imaging studies, including researchers from the Departments of Vascular Surgery and Pediatrics and Adolescent Medicine at University Hospital Erlangen, one of the leading centers for clinical \ac{OptA} research~\cite{Knieling2025-primer}. Requirements R1--R6 were derived from this collaboration, and the interface was revised iteratively in response to their feedback. This resulted in default workflows that closely align with clinical \ac{OPUS} study procedures. Given that structured training has been shown to improve \ac{MSOPUS} measurement reproducibility~\cite{Li2024-teachability}, OPTARI provides user-friendly documentation that also includes step-by-step guides and training videos. The tool is actively being used in ongoing clinical studies, with corresponding manuscripts currently in preparation.

\subsection{Performance of the Automated Analysis}\label{sect:evaluation-analysis}

The automated \ac{ROI} placement implemented in OPTARI's segmentation module is based on a muscle segmentation approach that was previously developed and validated~\cite{Schillinger2025-padseg}. In that work, a UNet-based convolutional neural network trained to delineate the target muscle on co-registered ultrasound achieved a Dice score of 0.86 and a 95th-percentile Hausdorff distance of 1.92\,mm across 5-fold cross-validation, outperforming Segment Anything Model (SAM)- and MedSAM-based alternatives. Classification of \ac{PAD} patients versus healthy volunteers based on the automatically generated \acp{ROI} was comparable to manual expert annotation, with an \ac{AUROC} of 0.76 (automatic) vs.\ 0.77 (manual).

Subsequent work further validated this approach on a post-exercise oxygenation recovery dataset, comparing automatically and physician-annotated \ac{MSOPUS}-derived oxygen saturation parameters~\cite{Buehler2026-msotpad}. Agreement between the two annotation methods was good to excellent across most summary parameters (e.g.\ mean recovery: intraclass correlation coefficient (ICC) 0.934 (95\% CI [0.834, 0.967]), concordance correlation coefficient (CCC) 0.933 (95\% CI [0.906, 0.953])), with near-zero bias. Classification performance for separating \ac{PAD} patients from healthy volunteers was likewise comparable between methods (e.g.\ peak \ac{AUROC} 0.966 (95\% CI [0.896, 0.997]) automatic vs.\ 0.981 (95\% CI [0.958, 1.000]) manual).

\begin{figure*}[htbp]
  \centering
  \includegraphics[width=0.9\textwidth]{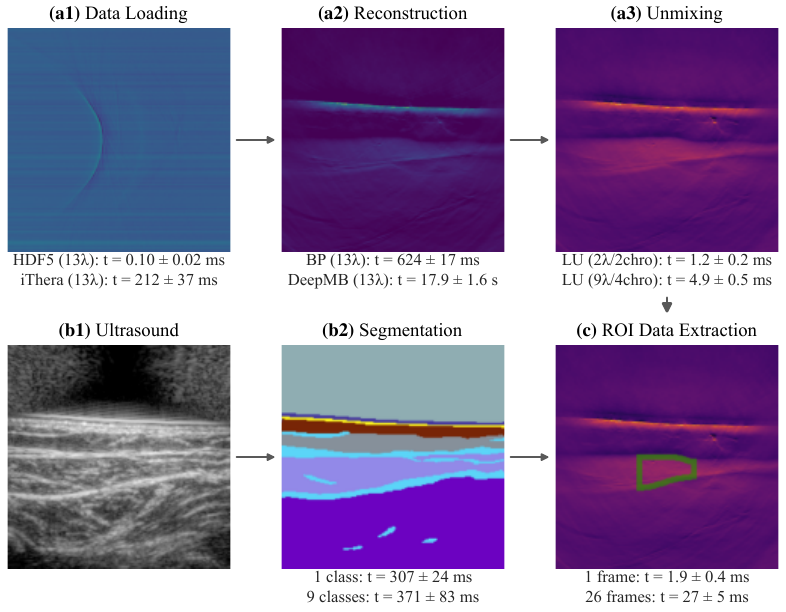}
\caption{\label{fig:workflow-timing} Automated batch-mode analysis applied to one representative scan (13 wavelengths, $\lambda$). Each panel reports two configuration choices for its step. Times are reported as mean $\pm$ standard deviation over 100 repetitions, measured on a standard laptop (Intel Core i7-1255U, 1.7\,GHz, 16\,GB RAM). HDF5 data are loaded lazily, so the read cost may fall into the following steps. The default configuration (HDF5 loading, \ac{BP}, two-chromophore \ac{LU}, single-class segmentation, single-frame extraction) took 0.93\,$\pm$\,0.03\,s in total.}
\end{figure*}

Figure~\ref{fig:workflow-timing} benchmarks the batch-mode pipeline on one representative scan (13 acquired wavelengths, each reconstructed into a 400$\times$400\,px image), comparing two configurations at each step while holding the rest of the pipeline fixed. Timings were measured on a standard machine (Intel Core i7-1255U, 1.7\,GHz, 16\,GB RAM), rather than high-performance hardware. With the default configuration (HDF5-loaded data, \ac{BP} reconstruction, two-chromophore \ac{LU}, single-class segmentation, single-frame \ac{ROI} extraction), the complete sequence took 0.93\,$\pm$\,0.03\,s per scan, of which reconstruction accounts for approximately 67\,\%,
whereas segmentation and automatic \ac{ROI} placement together contributed only 0.31\,$\pm$\,0.02\,s. For comparison, manual delineation of a single complex \ac{ROI} can take up to one minute, and a complete multi-class segmentation mask took 13.5\,$\pm$\,3.1\,min (10 timed annotations by an experienced operator). At study scale, the default configuration processes the dataset from \cite{Buehler2026-data} (1{,}977 scans from 102 subjects) in roughly 31\,min of unattended computation, excluding manual quality checks.

Together, these results suggest that automated segmentation is a viable alternative to manual \ac{ROI} annotation in \ac{MSOPUS}-based \ac{PAD} studies without a clear loss in group discrimination, while removing the operator-dependence of the annotation step and substantially reducing the required analysis time.

\section{Discussion}

To our knowledge, OPTARI is the only open-source tool that combines a graphical interface with the reconstruction, unmixing, and multi-patient workflow capabilities needed for large-scale clinical \ac{OPUS} studies. Rather than being developed from scratch, OPTARI builds on existing open-source infrastructure such as napari and \ac{PATATO}, adding a clinically oriented workflow and automated parameter selection.

OPTARI also aligns closely with the priorities identified in the \ac{IPASC} community's standardization roadmap~\cite{Assi2023-roadmap}. The consortium named automated \ac{ROI} delineation as a priority for overcoming segmentation and interpretation challenges, which is directly addressed by OPTARI's automated segmentation module (R4). Additionally, the roadmap for clinical adoption called for easy-to-use and openly available software for reconstruction and analysis, as well as for educational resources, both reflected by OPTARI's \ac{GUI} (R1, R2) and documentation (R5). Their roadmap for data management called for consolidating the \ac{IPASC} data format across research groups, which OPTARI supports by reading and writing \ac{IPASC} files and by including its metadata schema (R6).

Several limitations remain. OPTARI specifically targets clinical two-dimensional handheld \ac{OPUS} data and it does not support three-dimensional/volumetric scans or data from preclinical systems. The \ac{GUI} also lacks a true multi-viewer layout with synchronized annotations across layers, a limitation of napari's grid-view mode. Automated \ac{ROI} placement performance is further limited by the similarity between target data and the training distribution, and may require fine-tuning for new anatomical sites or scanner types. In practice, automatically placed \acp{ROI} should be reviewed rather than accepted unchecked, which reduces but does not eliminate the time saved.

OPTARI is research software intended for the analysis of data from clinical studies and has not undergone conformity assessment as a medical device. It is not intended for diagnosis or treatment decisions. The tool remains under active development, and we welcome further community contributions.

\subsection*{Disclosures}
V.N. is a founder and equity owner of iTheraNEXT Imaging GmbH, Maurus OY, sThesis GmbH, Spear UG, Biosense Innovations P.C. and I3 Inc. F.K. holds a patent on an \ac{OptA} imaging system together with iThera Medical GmbH. All other authors declare that they have no financial interests, commercial affiliations, or other potential conflicts of interest that could have influenced the objectivity of this research or the writing of this paper.

\subsection*{Code, Data, and Materials Availability}

The OPTARI source code and executables are available at \linkable{https://github.com/Mo-Sc/OPTARI} under the BSD-3-Clause license, with documentation at \linkable{https://mo-sc.github.io/OPTARI}. The analyses in this paper used version 0.8.0. The extended \ac{PATATO} version used by OPTARI is available at \linkable{https://github.com/Mo-Sc/patato}. Pretrained segmentation models are downloaded automatically on first use. The DeepMB adapter requires model weights, which are available from the original authors upon reasonable request~\cite{Dehner2023-deepmb}. The clinical \ac{PAD} dataset used in Section~\ref{sect:evaluation-analysis} is available on Zenodo~\cite{Buehler2026-data, Buehler2026-msotpad}.

\subsection*{Acknowledgments}

This research received funding from the Bavarian Ministry of Economic Affairs, Regional Development and Energy (StMWi) (DIK-2208-0014/ DIK0464/02, SENDERO, granted to K.B. and iThera Medical GmbH), from the Munich Institute of Robotics and Machine Intelligence (MIRMI) of the Technical University of Munich (TUM) (RoMSOT, granted to V.N.); from the Deutsche Forschungsgemeinschaft (DFG) - 455422993 as part of the Research Unit FOR 5298 (iMAGO, subproject TP2, GZ: NT 3/32-1, granted to V.N.); from the European Research Council (ERC) under the European Union’s Horizon 2020 research and innovation programme under grant agreement No 694968 (PREMSOT, granted to V.N.); and from the European Union’s Horizon Europe research and innovation programme under Grant Agreement No. 101291160 (QMSOT, granted to V.N.). 

LLM tools were used to assist with language refinement during manuscript preparation and to refactor portions of the OPTARI codebase. All content was reviewed and verified by the authors.

\bibliography{report}   %
\bibliographystyle{spiejour}   %

\end{spacing}
\end{document}